\documentclass[12pt]{article}
\RequirePackage[OT1]{fontenc}
\RequirePackage{amsthm,amsmath}
\RequirePackage[authoryear]{natbib}

\usepackage{footnote}
\usepackage{algorithm}
\usepackage{algpseudocode}
\usepackage{amsfonts}
\usepackage{color}

\usepackage{graphicx}
\usepackage{amsmath}
\usepackage{amssymb}
\usepackage{algorithm}
\newtheorem{theorem}{Theorem}
\newtheorem{lemma}{Lemma}
\newtheorem{example}{Example}

\begin{document}

\title{\large\bf Modeling Sparse Data Using MLE with Applications to Microbiome Data}
\author{Hani Aldirawi and Jie Yang\\
	California State University-San Bernardino and University of Illinois at Chicago}

\maketitle

\begin{abstract}
Modeling sparse data such as microbiome and  transcriptomics (RNA-seq) data is very challenging due to the exceeded number of zeros and skewness of the distribution. Many probabilistic models have been used for modeling sparse data, including Poisson, negative binomial, zero-inflated Poisson, and zero-inflated negative binomial models. One way to identify the most appropriate probabilistic models for zero-inflated or hurdle models is based on the p-value of the Kolmogorov-Smirnov (KS) test. The main challenge for identifying the probabilistic model is that the model parameters are typically unknown in practice. This paper derives the maximum likelihood estimator (MLE) for a general class of zero-inflated and hurdle models. We also derive the corresponding Fisher information matrices for exploring the estimator's asymptotic properties. We include new probabilistic models such as zero-inflated beta binomial and zero-inflated beta negative binomial models. Our application to microbiome data shows that our new models are more appropriate for modeling microbiome data than commonly used models in the literature. 
\end{abstract}


\section{Introduction}\label{sec1}

The microbiome, a dynamic ecosystem of microorganisms (bacteria, archaea, fungi, and viruses) that live in and on us, plays a vital role in host-immune responses resulting in significant effects on host health (see, for example, \cite{metwally2018review}). Dysbiosis of the microbiome has been linked to diseases including asthma, obesity, diabetes, transplant rejection, and inflammatory bowel disease~\citep{Vatanen2016VariationHumans,pflughoeft2012human,Cho2012TheDisease,rani2016diverse}. 
These observations suggest that modulation of the microbiome could become an important therapeutic modality for some diseases~\citep{ sehrawat2021probiotics, gupta2020therapies, Yatsunenko2012HumanGeography}. 

Modeling microbiome data is very challenging due to the exceeding number of zeros in the data~\citep{gloor2017microbiome, knights2011supervised}. Dealing with zeros is one of the biggest challenges in microbiome and transcriptomics studies \citep{xia2018statistical}. It is challenging to model those features which are skewed and zero-inflated \citep{chen2012statistical}. Table~\ref{countTableExample} is a toy example of taxonomic profile with a dimension of $3\times 6$, where $3$ denotes the number of microbial features and $6$ denotes the number of metagenomic samples. The table shows the sparsity of the mirobiome data. Therefore, zero-inflated Poisson (ZIP), zero-inflated negative binomial (ZINB), Poisson hurdle (PH), and negative binomial hurdle (NBH) models are commonly used to model microbiome data \citep{metwally2018review}.
\begin{table}[]
\caption{A Toy Example of Taxonomic Profile Count Table}
\label{countTableExample}
\begin{center}
    \begin{tabular}{||c| c c c c c c || c||} 
			\hline
			Species/Sample & $S_1$ & $S_2$ &$S_3$ &$S_4$ &$S_5$ &$S_6$ &Total \\ [0.5ex] 
			\hline\hline
			$\textit{Streptococcus pneumoniae}$ & $0$ & 0& $102$ &0 & $3$ & 0 & $105$ \\ 
			\hline
			$\textit{Escherichia coli}$ & $13$&0 & 0 &75 & $0$ & $0$ & $88$ \\
			\hline
			$\textit{Staphylococcus aureus}$ &  0&$14$& 0 & $0$ &138 & $0$& $152$ \\ 
			\hline\hline
			Total& $13$ & $14$& 102 & $75$ & 141& $0$ &345 \\
			\hline
		\end{tabular}\vspace{1em}
\end{center}
\end{table}

The  selection  of  an  appropriate  probabilistic  model  is critical for microbiome studies. For example, in order to determine if  there  is  an  association  between  a microbiome  feature (such  as a bacteria), and the disease, we may need to detect the significance of the difference between two groups of records. With appropriate probabilistic models identified successfully, we can improve the power of the statistical test significantly.
Recently, \cite{aldirawi2019identifying} proposed a statistical procedure for identifying the most appropriate discrete probabilistic models for zero-inflated or hurdle models based on the p-value of the discrete Kolmogorov-Smirnov (KS) test. The same procedure could be used for more general zero-inflated or hurdle models, including the ones with continuous baseline distributions. More specifically, the goal is to test if the sample ${\mathbf X} = \{ X_{1},X_{2},..,X_{n}  \}$ comes from a discrete or mixed distribution with cumulative distribution function (CDF) $F_{\boldsymbol\theta}(x)$ where the parameter(s) $\boldsymbol\theta$ is unknown. Algorithm~1, which is regenerated from \cite{aldirawi2019identifying}, provides our procedures in details.

\begin{algorithm}
\caption{Estimating p-value of KS test}
\begin{algorithmic}[1]
\State \emph{Given ${\mathbf X}=(X_1, X_2, \cdots X_n$)}
\State \emph{For $b = 1, \ldots, B$,
resample $X$ with replacement to get a bootstrapped sample ${\mathbf X}^{(b)} = \{X^{(b)}_{1}, \cdots, X_{n}^{(b)}\}$.}
\State \emph{For each b, calculate the MLE $\hat{\boldsymbol\theta}^{(b)}$ of $\boldsymbol\theta$.}
\State \emph{Simulate ${\mathbf X}^{(c)} = \{X^{(c)}_1, \ldots, X^{(c)}_n\}$ iid from $F_{\hat{\boldsymbol\theta}^{(b)}}$, which is the CDF $F_{\boldsymbol\theta}(x)$ with parameter $\boldsymbol\theta = \hat{\boldsymbol\theta}^{(b)}$. }
\State \emph{Calculate the KS statistic $D_{n}^{(b)}={\rm sup}_{x}\lvert \hat F^{(c)}_{n}(x)- F_{\hat{\boldsymbol\theta}^{(b)}}(x)\rvert$, where $\hat F^{(c)}_{n}(x)$ is the empirical distribution function of ${\mathbf X}^{(c)}$. }
\State \emph{Estimate the p-value by 
$\frac{\#\{b \mid D_{n}^{(b)} > D_{n}\}+1}{B+1}$
where $D_{n} = {\rm sup}_{x}\lvert \hat F_{n}(x)- F_{\hat{\boldsymbol\theta}}(x)\rvert$ is the KS statistic based on the original data and its MLE $\hat{\boldsymbol\theta}$.}
\end{algorithmic}
\end{algorithm}

Although the following procedure and algorithm were described, their theoretical justifications were not  provided in ~\cite{aldirawi2019identifying}. One major step in the above algorithm is to estimate the distribution parameters (step 3) using the maximum likelihood estimate (MLE) method. In this paper, we develop a general MLE procedure for estimating the parameters for general zero-inflated and hurdle models. In addition, we discuss the asymptotic properties and Fisher information matrix for MLEs, which can be used for building up the confidence intervals of the distribution parameters. For modeling microbiome data, we recommend zero-inflated (or hurdle) beta binomial or beta negative binomial models.

\section{Zero-altered or hurdle models and their MLEs}\label{sec:hurdle}

Zero-altered models, also known as {\it hurdle models}, have been used for modeling data with an excess or deficit of zeros (see, for example, ~\cite{metwally2018review}, for a review). A general hurdle model consists of two components, one generating the zeros and the other generating non-zeros (or positive values for many applications). Given a baseline distribution $f_{\boldsymbol\theta}(y)$, which could be the probability mass function (pmf) of a discrete distribution or the probability density function (pdf) of a condinuous distribution, with parameter(s) $\boldsymbol\theta = (\theta_1, \ldots, \theta_p)^T$,  the distribution function of the corresponding hurdle model can be written as follows:
\begin{equation}\label{eq:hurdle}
f_{\rm ZA}(y\mid \phi, \boldsymbol\theta) = \phi {\mathbf 1}_{\{y=0\}} + (1-\phi) f_{\rm tr}(y\mid \boldsymbol\theta) {\mathbf 1}_{\{y\neq 0\}}
\end{equation}
where $\phi \in [0, 1]$ is the weight parameter of zeros, $f_{\rm tr}(y\mid \boldsymbol\theta) = [1-p_0(\boldsymbol\theta)]^{-1} f_{\boldsymbol\theta} (y), y\neq 0$ is the pmf or pdf of the zero-truncated baseline distribution, and $p_0(\boldsymbol\theta) = f_{\boldsymbol\theta}(0)$ for discrete baseline distributions or simply $0$ for continuous baseline distributions.
Examples with discrete baseline distributions include zero-altered Poisson (ZAP) or Poisson hurdle (PH), zero-altered negative binomial (ZANB) or negative binomial hurdle (NBH) models and others, where model~\eqref{eq:hurdle} provides a new pmf. Examples with continuous basedline distributions include zero-altered Gaussian (ZAG) or Gaussian hurdle (GH), zero-altered lognormal or lognormal hurdle models and others, where model~\eqref{eq:hurdle} is indeed a mixture distribution with a discrete part with a probability mass $\phi$ at $[Y=0]$ and a continuous component in $[Y \neq 0]$ with density function $(1-\phi) f_{\boldsymbol\theta}(y), y\neq 0$.

The zero-altered models can actually be defined with a fairly general baseline distribution equiped with a cumulative distribution function (cdf) $F_{\boldsymbol\theta}(y) = P_{\boldsymbol\theta} (Y\leq y)$. Its corresponding cdf is defined as follows:
\[
F_{\rm ZA}(y\mid \phi, \boldsymbol\theta) = P_{\rm ZA}(Y \leq y\mid \phi, \boldsymbol\theta) = \phi {\mathbf 1}_{\{y \geq 0\}} + (1-\phi) F_{\rm tr}(y\mid \boldsymbol\theta) 
\]
where $F_{\rm tr}(y\mid \boldsymbol\theta) = [F_{\boldsymbol\theta}(y) - P_{\boldsymbol\theta}(Y=0) {\mathbf 1}_{\{y \geq 0\}}]/[1-P_{\boldsymbol\theta}(Y=0)]$ is a zero-truncated cdf. 
In this paper, we assume that the baseline distribution is either discrete or continuous with pmf or pdf $f_{\boldsymbol\theta}(y)$.

\subsection{Maximum likelihood estimate for zero-altered or hurdle model}

The parameters of hurdle model~\eqref{eq:hurdle} include both $\phi$ and $\boldsymbol\theta$. Let $Y_1, \ldots, Y_n$ be a random sample from model~\eqref{eq:hurdle}. 
The likelihood function of $(\phi, \boldsymbol\theta)$ is
\begin{equation}\label{eq:hurdelmle}
L(\phi, \boldsymbol\theta) = \phi^{n-m} (1-\phi)^m \cdot \prod_{i:Y_i\neq 0} f_{\rm tr}(Y_i\mid \boldsymbol\theta)
\end{equation}
where $m=\#\{i:Y_i\neq 0\}$ is the number of nonzero observations. Since $\phi$ and $\boldsymbol\theta$ are separable in the likelihood function, we obtain the following theorem.

\begin{theorem}\label{thm:hurdlemle}
For model~\eqref{eq:hurdle} with zero-truncated pmf or pdf $f_{\rm tr}(y\mid \boldsymbol\theta)$, the maximum likelihood estimate (MLE) maximizing \eqref{eq:hurdelmle} is
\[
\hat{\phi}=1-\frac{m}{n},\>\>\> 
\hat{\boldsymbol\theta}={\rm argmax}_{\boldsymbol\theta} \prod_{i:Y_i\neq 0} f_{\rm tr}(Y_i\mid \boldsymbol\theta)
\]
\end{theorem}
Recall that $f_{\rm tr}(y\mid \boldsymbol\theta) = [1-p_0(\boldsymbol\theta)]^{-1} f_{\boldsymbol\theta}(y), y\neq 0$ for discrete baseline functions or simply $f_{\boldsymbol\theta}(y), y\neq 0$ for continuous baseline functions.

\begin{example}\label{ex:ZABB}{
For zero-altered beta binomial or beta binomial hurdle (BBH) distribution, the pmf of the baseline distribution is 
\[
f_{\boldsymbol\theta} (y) = \dbinom{n}{y} \frac{{ \rm beta}(y + \alpha, n - y + \beta)}{{ \rm beta}(\alpha,\beta)}
\]
where $\boldsymbol\theta = (n, \alpha, \beta)$, $y=0, 1, \ldots, n$ and
\begin{equation*}
p_0(\boldsymbol\theta) = \frac{\Gamma(n + \beta) \Gamma(\alpha + \beta)}{\Gamma(n + \alpha + \beta)  \Gamma(\beta)}
\end{equation*}
	Let $L(\boldsymbol\theta)$ be the likelihood of zero-truncated beta binomial distribution, then 
\begin{eqnarray*}
	L(n,\alpha,\beta)&=& {argmax}_{\boldsymbol\theta}  \frac{\prod_{i: y_i \neq 0}f_{\theta}(y_i)}{[1-p_0(\boldsymbol\theta)]^m}=\left(\frac{\Gamma(\alpha+n+\beta)\Gamma(\beta)}{\Gamma(\alpha+n+\beta)\Gamma(\beta)-\Gamma(n+\beta)\Gamma(\alpha+\beta)}\right)^{m}\cdot\\	
	& &\prod_{i=1}^{m}   \left(\frac{\Gamma(n+1)\Gamma(y_i+\alpha)\Gamma(n-y_i+\beta)
		\Gamma(\alpha+\beta)}{\Gamma(y_{i}+1)\Gamma(n-y_i+1)\Gamma(\alpha+n+\beta)\Gamma(\alpha) \Gamma(\beta)}\right)  
\end{eqnarray*}
	
 The loglikelihood of zero-truncated beta binomial is given by:
\begin{eqnarray*}  
l(n,\alpha,\beta)&=& m\log\Gamma(n+1)+m\log\Gamma(\alpha+\beta)-m\log\Gamma(\alpha)+\sum_{i=1}^{n}\log \Gamma(y_{i}+\alpha) \\ 
&-& m\log\left(\Gamma(\alpha+n+\beta)\Gamma(\beta)-\Gamma(n+\beta)\Gamma(\alpha+\beta)\right)+\sum_{i=1}^{m}\log \Gamma(n-y_{i}+\beta)\\ &-& \sum_{i=1}^{m}\log \Gamma(y_{i}+1) -\sum_{i=1}^{m}\log \Gamma(n-y_{i}+1)
\end{eqnarray*}

Let $\Psi(\cdot) = \Gamma'(\cdot)/\Gamma(\cdot)$, known as the {\it digamma} function. In order to calculate the MLE, the following formulae are needed:
\begin{eqnarray*}
	 \frac{\partial l(n,\alpha,\beta)}{\partial n}& =& m\left(\frac{\exp(\log B-\log A)(\psi(n+\beta)-\psi(n+\alpha+\beta))}{1-\exp(\log B-\log A)}\right)+m\psi(n+1)\\
	 &&+
\sum_{i=1}^{m} \psi(n-y_{i}+\beta) -\sum_{i=1}^{m} \psi(n-y_{i}+1)-m\psi(\alpha+n+\beta) \\
	 \frac{\partial l(n,\alpha,\beta)}{\partial \alpha} &=& m\left(\frac{\exp(\log B-\log A)(\psi(n+\beta)-\psi(n+\alpha+\beta))}{1-\exp(\log B-\log A)}\right)\\
	 &&+\sum_{i=1}^{m} \psi(y_{i}+\alpha) +m\psi(\alpha+\beta)-m\psi(\alpha+n+\beta)-m\psi(\alpha) \\
	 \frac{\partial l(n,\alpha,\beta)}{\partial \beta} &=& m\left(\frac{\exp(\log B-\log A)(\psi(n+\beta)+\psi(\alpha+\beta)-\psi(\alpha+n+\beta)-\psi(\beta))}{1-\exp(\log B-\log A)}\right)\\
	 &&-m\psi(\beta)+\sum_{i=1}^{m} \psi(n-y_{i}+\beta) +m\psi(\alpha+\beta)-m\psi(\alpha+n+\beta) 
\end{eqnarray*}
where $A=\Gamma(n+\alpha+\beta) \Gamma(\beta)$, and $B=\Gamma(n+\beta) \Gamma(\alpha+\beta)$.
}\hfill{$\Box$}
\end{example}

\subsection{Asymptotic properties and Fisher information matrix of hurdle MLEs}

Let $Y_1, \ldots, Y_n$ be a random sample from hurdle model~\eqref{eq:hurdle}. In order to find the MLE numerically, we often consider the log-likelihood function
\begin{eqnarray*}
l(\phi, \boldsymbol\theta) = \log L(\phi, \boldsymbol\theta) &&= \log\phi \cdot \sum_{i=1}^n {\mathbf 1}_{\{Y_i=0\}} + \log(1-\phi) \cdot \sum_{i=1}^n {\mathbf 1}_{\{Y_i\neq 0\}}\\
&&-\ \log [1-p_0(\boldsymbol\theta)] \sum_{i=1}^n {\mathbf 1}_{\{Y_i\neq 0\}} + \sum_{i=1}^n \log f_{\boldsymbol\theta}(Y_i) {\mathbf 1}_{\{Y_i\neq 0\}}
\end{eqnarray*}
whose first derivatives are
\begin{eqnarray*}
\frac{\partial l}{\partial\phi} &=& \frac{1}{\phi} \cdot \sum_{i=1}^n {\mathbf 1}_{\{Y_i=0\}} - \frac{1}{1-\phi} \cdot \sum_{i=1}^n {\mathbf 1}_{\{Y_i\neq 0\}}\\
\frac{\partial l}{\partial \boldsymbol\theta} &=& \frac{p_0(\boldsymbol\theta)}{1-p_0(\boldsymbol\theta)}\cdot \frac{\partial \log p_0(\boldsymbol\theta)}{\partial \boldsymbol\theta} \sum_{i=1}^n {\mathbf 1}_{\{Y_i\neq 0\}} + \sum_{i=1}^n \frac{\partial \log f_{\boldsymbol\theta}(Y_i)}{\partial \boldsymbol\theta} {\mathbf 1}_{\{Y_i\neq 0\}}
\end{eqnarray*}

\begin{lemma}\label{lem:hurdelmlefirst}
\[
E\left(\frac{\partial l}{\partial \phi}\right) = 0 \>\>\mbox{ and }\>\> E\left(\frac{\partial l}{\partial \boldsymbol\theta}\right) = \frac{n(1-\phi)}{1-p_0(\boldsymbol\theta)} \cdot E\left[\frac{\partial \log f_{\boldsymbol\theta}(Y')}{\partial \boldsymbol\theta} \right]
\]
where $Y'$ follows the baseline distribution $f_{\boldsymbol\theta}(y)$.
\end{lemma}

\medskip\noindent
{\bf Proof of Lemma~\ref{lem:hurdelmlefirst}:} 
Since $P(Y_i=0) = \phi$, then $E(\partial l/\partial\phi) = 0$. Let $Y_1', \ldots, Y_n'$ be iid $\sim f_{\boldsymbol\theta} (y)$. Then
\begin{eqnarray*}
E\left[\frac{\partial \log f_{\boldsymbol\theta}(Y_i)}{\partial \boldsymbol\theta} {\mathbf 1}_{\{Y_i\neq 0\}}\right] \ &&=\ \frac{1-\phi}{1-p_0(\boldsymbol\theta)} \cdot E\left[\frac{\partial \log f_{\boldsymbol\theta}(Y_i')}{\partial \boldsymbol\theta} {\mathbf 1}_{\{Y_i'\neq 0\}}\right]  \\
&&= \frac{1-\phi}{1-p_0(\boldsymbol\theta)} \cdot \left\{ E\left[\frac{\partial \log f_{\boldsymbol\theta}(Y_i')}{\partial \boldsymbol\theta} \right] - p_0(\boldsymbol\theta)\cdot \frac{\partial \log p_0(\boldsymbol\theta)}{\partial \boldsymbol\theta}\right\}
\end{eqnarray*}
Then
\[
E\left(\frac{\partial l}{\partial \boldsymbol\theta}\right) = \frac{1-\phi}{1-p_0(\boldsymbol\theta)} \cdot \sum_{i=1}^n E\left[\frac{\partial \log f_{\boldsymbol\theta}(Y_i')}{\partial \boldsymbol\theta} \right] = \frac{n(1-\phi)}{1-p_0(\boldsymbol\theta)} \cdot E\left[\frac{\partial \log f_{\boldsymbol\theta}(Y')}{\partial \boldsymbol\theta} \right]
\]
\hfill{$\Box$}

As a direct corollary of Theorem~17 in \cite{ferguson1996}, the MLEs of hurdle model have strong consistency under fairly general conditions.

\begin{theorem}\label{thm:hurdlemleconsistency}
Let $Y_1, \ldots, Y_n$ be a random sample from hurdle model~\eqref{eq:hurdle} with true parameter value $(\phi_0, \boldsymbol\theta_0) \in (0,1)\times \boldsymbol\Theta$, where $\boldsymbol\Theta$ is compact. Let
$\hat{\phi}=n^{-1} \sum_{i=1}^n {\mathbf 1}_{\{Y_i =0\}}$ and $\hat{\boldsymbol\theta}={\rm argmax}_{\boldsymbol\theta \in \boldsymbol\Theta} \prod_{i:Y_i\neq 0} f_{\rm tr}(Y_i\mid \boldsymbol\theta)$ be the MLEs.
Suppose (1) $f_{\boldsymbol\theta}(y)$ is continuous in $\boldsymbol\theta$ for all $y$; (2) $f_{\boldsymbol\theta}(y) = f_{\boldsymbol\theta_0}(y)$ for all $y$ always implies $\boldsymbol\theta = \boldsymbol\theta_0$; and (3) there exists a nonnegative function $K(y)$ such that $E[K(Y)] < \infty$ for $Y\sim f_{\rm tr}(y; \boldsymbol\theta_0)$ and $\log[f_{\rm tr}(y\mid \boldsymbol\theta)/f_{\rm tr}(y \mid \boldsymbol\theta_0)] \leq K(y)$ for all $y\neq 0$ and $\boldsymbol\theta \in \boldsymbol\Theta$. Then $\hat\phi \stackrel{\rm a.s.}{\longrightarrow} \phi_0$ and $\hat{\boldsymbol\theta}_n \stackrel{\rm a.s.}{\longrightarrow} \boldsymbol\theta_0$ as $n$ goes to infinity.
\end{theorem}

Under regularity conditions, see for example, Chapter 18 in \cite{ferguson1996} or Section~5f in \cite{rao1973linear}, $E[\partial \log f_{\boldsymbol\theta}(Y')/\partial \boldsymbol\theta]=0$ and thus $E\left(\partial l/\partial \boldsymbol\theta\right) =0$ according to Lemma~\ref{lem:hurdelmlefirst}. We can further calculate the Fisher information matrix of the random sample
\begin{equation}\label{eq:fisher_information}
{\mathbf F}(\phi, \boldsymbol\theta) = 
-\left[
\begin{array}{cc}
E\left(\frac{\partial^2 l}{\partial \phi^2}\right) & E\left(\frac{\partial^2 l}{\partial \phi \partial \boldsymbol\theta^T}\right) \\
E\left(\frac{\partial^2 l}{\partial\boldsymbol\theta \partial \phi}\right) & E\left(\frac{\partial^2 l}{\partial \boldsymbol\theta \partial\boldsymbol\theta^T}\right)
\end{array}\right]
\end{equation}

\begin{theorem}\label{thm:hurdle_Fisher}
Let $Y_1, \ldots, Y_n$ be a random sample from hurdle model~\eqref{eq:hurdle}. Under regularity conditions, the Fisher information matrix of the sample is
\[
{\mathbf F}_{\rm ZA} = 
n \left[\begin{array}{cc}
\phi^{-1} (1-\phi)^{-1} & {\mathbf 0}^T\\
{\mathbf 0} & {\mathbf F}_{\rm ZA22}
\end{array}\right]
\]
where 
\[
{\mathbf F}_{\rm ZA22} = -\frac{1-\phi}{1-p_0(\boldsymbol\theta)} \left(E\left[\frac{\partial^2 \log f_{\boldsymbol\theta}(Y')}{\partial \boldsymbol\theta \partial \boldsymbol\theta^T} \right] + \frac{p_0(\boldsymbol\theta)}{1-p_0(\boldsymbol\theta)} \cdot \frac{\partial \log p_0(\boldsymbol\theta)}{\partial \boldsymbol\theta}\cdot \frac{\partial \log p_0(\boldsymbol\theta)}{\partial \boldsymbol\theta^T}\right)
\]
and $Y'$ follows the baseline distribution $f_{\boldsymbol\theta}(y)$.
\end{theorem}

\medskip\noindent
{\bf Proof of Theorem~\ref{thm:hurdle_Fisher}:}
Since 
\begin{eqnarray*}
\frac{\partial\log f_{\rm ZA} (y \mid \phi, \boldsymbol\theta)}{\partial \phi} &=& \phi^{-1} {\mathbf 1}_{\{y=0\}} - (1-\phi)^{-1} {\mathbf 1}_{\{y\neq 0\}}\\
\frac{\partial\log f_{\rm ZA} (y \mid \phi, \boldsymbol\theta)}{\partial \boldsymbol\theta} &=& \frac{p_0(\boldsymbol\theta)}{1-p_0(\boldsymbol\theta)} \cdot \frac{\partial\log p_0(\boldsymbol\theta)}{\partial \boldsymbol\theta} {\mathbf 1}_{\{y\neq 0\}} + \frac{\partial \log f_{\boldsymbol\theta}(y)}{\partial \boldsymbol\theta} {\mathbf 1}_{\{y\neq 0\}}
\end{eqnarray*}
Then 
\begin{eqnarray*}
\frac{\partial^2\log f_{\rm ZA} (y \mid \phi, \boldsymbol\theta)}{\partial \phi^2} &=& -\phi^{-2} {\mathbf 1}_{\{y=0\}} - (1-\phi)^{-2} {\mathbf 1}_{\{y\neq 0\}}\\
\frac{\partial^2\log f_{\rm ZA} (y \mid \phi, \boldsymbol\theta)}{\partial \phi \partial \boldsymbol\theta^T} &=& {\mathbf 0}^T\\
\frac{\partial^2\log f_{\rm ZA} (y \mid \phi, \boldsymbol\theta)}{\partial \boldsymbol\theta \partial \phi} &=& {\mathbf 0}\\
\frac{\partial^2\log f_{\rm ZA} (y \mid \phi, \boldsymbol\theta)}{\partial \boldsymbol\theta \partial \boldsymbol\theta^T} &=&  \frac{p_0(\boldsymbol\theta)}{[1-p_0(\boldsymbol\theta)]^2} \cdot \frac{\partial\log p_0(\boldsymbol\theta)}{\partial \boldsymbol\theta} \cdot \frac{\partial\log p_0(\boldsymbol\theta)}{\partial \boldsymbol\theta^T}{\mathbf 1}_{\{y\neq 0\}}\\
&+&\frac{p_0(\boldsymbol\theta)}{1-p_0(\boldsymbol\theta)} \cdot \frac{\partial^2\log p_0(\boldsymbol\theta)}{\partial \boldsymbol\theta \partial \boldsymbol\theta^T} {\mathbf 1}_{\{y\neq 0\}} +\frac{\partial^2 \log f_{\boldsymbol\theta}(y)}{\partial \boldsymbol\theta \partial \boldsymbol\theta^T} {\mathbf 1}_{\{y\neq 0\}}
\end{eqnarray*}
The rest of the conclusions can be obtained via $l(\phi, \boldsymbol\theta) = \sum_{i=1}^n \log f_{\rm ZA} (Y_i \mid \phi, \boldsymbol\theta)$.
\hfill{$\Box$}

As direct conclusions of Theorem~18 in \cite{ferguson1996}: (i) $\sqrt{n}(\hat\phi - \phi_0)  \stackrel{{\cal L}}{\rightarrow} N(0, \phi_0 (1-\phi_0))$; (ii) $\sqrt{n}(\hat{\boldsymbol\theta} - \boldsymbol\theta_0)  \stackrel{{\cal L}}{\rightarrow} N\left(0, {\mathbf F}_{\rm ZA22}^{-1} \right)$; and (iii) $\hat\phi$ and $\hat{\boldsymbol\theta}$ are asymptotically independent. These results can be used for building up confidence intervals of $\phi$ and $\boldsymbol\theta$.

\begin{example}\label{ex:ZAP}{\rm
For zero-altered Poisson or Poisson hurdle  distribution, the pmf of the baseline distribution is $f_\lambda(y)= e^{-\lambda}\lambda^{y}/y!$ with $p_0(\lambda) = e^{-\lambda}$. It can be verified that 
\[
E\left(\frac{\partial\log f_\lambda(Y')}{\partial \lambda}\right) = 0
\]
if $Y' \sim f_\lambda (y)$. The truncated pmf is
\[
\rm f_{\rm tr}(y \mid \lambda) = \frac{e^{-\lambda}}{1-e^{-\lambda}} \cdot \frac{\lambda^y}{y!}, \>\>\> y=1,2,\ldots 
\]
The loglikelihood for the zero-truncated Poisson is
\[
l(\lambda) = -m\lambda-m\log(1-e^{-\lambda}) + \sum_{i:Y_i>0} Y_i \cdot \log\lambda - \log(\prod_{i:Y_i>0} Y_i!)
\] 
The MLE $\hat\lambda$ of $\lambda$ solves the likelihood equation $\lambda = \bar{Y} (1-e^{-\lambda})$ with $\bar{Y} = m^{-1} \sum_{i:Y_i>0} Y_i$~, which can be solved numerically. If the true value $\lambda_0 \in [\lambda_1, \lambda_2]$ for some $0 < \lambda_1 < \lambda_2 < \infty$, then $K(y)$ in Theorem~\ref{thm:hurdlemle} can be chosen as
\[
K(y) = \log\frac{\lambda_2}{\lambda_1} \cdot y + \log\frac{1-e^{-\lambda_2}}{1-e^{-\lambda_1}} + \lambda_2 - \lambda_1
\]
Since there is no difference in practice as long as $0 < \lambda_1 < \hat\lambda < \lambda_2 < \infty$, we know $\hat\lambda \stackrel{\rm a.s.}{\rightarrow} \lambda_0$ as $n$ goes to infinity.

According to Theorem~\ref{thm:hurdle_Fisher}, the Fisher information matrix of the Poisson hurdle sample is
\[
{\mathbf F}_{\rm PH} = \left[\begin{array}{cc}
\frac{1}{\phi(1-\phi)} & 0 \\
0 & \frac{1-\phi}{1-e^{-\lambda}} \left(\frac{1}{\lambda} - \frac{e^{-\lambda}}{1-e^{-\lambda}}\right)
\end{array}\right]
\]
Note that $\frac{1}{\lambda} - \frac{e^{-\lambda}}{1-e^{-\lambda}} > 0$ as long as $\lambda > 0$.
}\hfill{$\Box$}
\end{example}


\begin{example}\label{ex:ZANB}{\rm
For zero-altered negative binomial or negative binomial hurdle distribution, the pmf of the baseline distribution with parameters $\boldsymbol\theta = (r, p) \in (0, \infty)\times [0,1]$ is given by
$f_{\boldsymbol\theta}(y)= \frac{\Gamma(y+r)}{\Gamma(y+1) \Gamma(r)} p^{y}(1-p)^{r}$, $y \in \{0, 1, 2, \ldots\}$. Then $p_0(\boldsymbol\theta) = (1-p)^r$. In order to apply Lemma~\ref{lem:hurdelmlefirst}, we obtain
\begin{eqnarray*}
\log f_{\boldsymbol\theta}(y) &=& \log\Gamma(y+r) - \log\Gamma(y+1) - \log\Gamma(r) + y\log p + r\log(1-p)\\
\frac{\partial\log f_{\boldsymbol\theta}(y)}{\partial r} &=& \Psi(y+r)-\Psi(r)+\log(1-p)\\
\frac{\partial\log f_{\boldsymbol\theta}(y)}{\partial p} &=& \frac{y}{p}-\frac{r}{1-p}
\end{eqnarray*}
where $\Psi(\cdot) = \Gamma'(\cdot)/\Gamma(\cdot)$ is known as the {\it digamma} function.

If $Y' \sim f_{\boldsymbol\theta}(y)$, then $E(Y') = pr/(1-p)$ and $E\left(\partial\log f_{\boldsymbol\theta}(Y')/\partial p\right) = 0$. On the other hand,
since $\Gamma(y) = \int_0^\infty t^{y-1} e^{-t} dt$ and $\Gamma'(y) = \int_0^\infty t^{y-1} e^{-t} \log t dt$ for $y>0$, then 

\begin{eqnarray*}
E\left(\Psi(Y'+r)\right) &=& \sum_{y=0}^\infty \frac{\Gamma'(y+r)}{\Gamma(y+r)}\cdot \frac{\Gamma(y+r)}{\Gamma(y+1)\Gamma(r)}p^y (1-p)^r\\
&=& \frac{(1-p)^r}{\Gamma(r)} \sum_{y=0}^\infty \frac{p^y}{y!} \Gamma'(y+r)\\
&=& \frac{(1-p)^r}{\Gamma(r)} \sum_{y=0}^\infty \frac{p^y}{y!} \int_0^\infty t^{y+r-1} e^{-t}\log tdt\\
&=& \frac{(1-p)^r}{\Gamma(r)} \int_0^\infty \left(\sum_{y=0}^\infty \frac{(pt)^y}{y!} e^{-pt}\right) \cdot t^{r-1} e^{-t(1-p)} \log t dt\\ 	
&=& \frac{(1-p)^r}{\Gamma(r)} \int_0^\infty t^{r-1} e^{-t(1-p)} \log t dt\>\>\> (\mbox{let }s=(1-p)t)\\ 	
&=& \frac{(1-p)^r}{\Gamma(r)} \int_0^\infty s^{r-1} e^{-s}\left[\log s - \log(1-p)\right] ds \cdot (1-p)^{-r}\\	
&=& \frac{1}{\Gamma(r)} \left[\int_0^\infty s^{r-1} e^{-s}\log s ds - \int_0^\infty s^{r-1} e^{-s} \log(1-p) ds\right]\\
&=& \frac{1}{\Gamma(r)} \left[\Gamma'(r) - \log(1-p) \Gamma(r)\right]\\
&=& \Psi(r) - \log(1-p)
\end{eqnarray*}
Therefore, $E\left(\partial\log f_{\boldsymbol\theta}(Y')/\partial r\right) =  E\left(\Psi(Y'+r)\right) - \Psi(r)+\log(1-p) = 0$.

According to Theorem~\ref{thm:hurdle_Fisher}, the Fisher information matrix of a random sample from the negative binomial hurdle distribution is
\[
{\mathbf F}_{\rm NBH} = \left[\begin{array}{cc}
\frac{1}{\phi(1-\phi)} & {\mathbf 0}^T\\
{\mathbf 0} & {\mathbf F}_{\rm NBH22}
\end{array}\right]
\]
with
\[
{\mathbf F}_{\rm NBH22} = 
\frac{1-\phi}{1-(1-p)^r} \left\{
\left[\begin{array}{cc}
A(y,r) & \frac{1}{1-p}\\
\frac{1}{1-p} & \frac{r}{p(1-p)^2}
\end{array}\right] -
\frac{(1-p)^r}{1-(1-p)^r}
\left[\begin{array}{cc}
\log^2(1-p) & -\frac{r\log (1-p)}{1-p}\\
-\frac{r\log(1-p)}{1-p} & \frac{r^2}{(1-p)^2}
\end{array}\right]\right\}
\]
$A(y,r)=\Psi_1(r) - E\Psi_1(Y' + r)$
where $\Psi_1(\cdot) = \Psi'(\cdot)$ is known as the {\it trigamma} function. Since $E\Psi_1(Y' + r)$ does not have a simple form for computation, a numerical solution for estimating it has been proposed by \cite{guo2020}. 
}\hfill{$\Box$}
\end{example}

\section{Zero-inflated models and their MLEs}\label{sec:zimodel}

Unlike zero-altered models, a zero-inflated model always assumes an excess of zeros. Besides zeros coming from the baseline distribution, such as Poisson or negative binomial, there are additional zeros modeled by a weight parameter $\phi \in [0,1]$. 

When the baseline distribution is discrete with a pmf $f_{\boldsymbol\theta}(y)$, the corresponding zero-inflated model has a pmf $f_{\rm ZI}(y\mid \phi, {\boldsymbol\theta}) = \phi {\mathbf 1}_{\{y=0\}} + (1-\phi) f_{{\boldsymbol\theta}}(y)
$ as well. In order to cover both pmf and pdf, we would rather write the distribution function of the zero-inflated distribution as
\begin{equation}\label{eq:zimodel}
f_{\rm ZI}(y\mid \phi, {\boldsymbol\theta}) = [\phi + (1-\phi) p_0(\boldsymbol\theta)] {\mathbf 1}_{\{y=0\}} + (1-\phi) f_{{\boldsymbol\theta}}(y) {\mathbf 1}_{\{y\neq 0\}}
\end{equation}
Recall that $p_0(\boldsymbol\theta) = f_{\boldsymbol\theta}(0)$ for pmf and $0$ for pdf. Similar to zero-altered models, when the baseline function $f_{\boldsymbol\theta}(y)$ is a pdf, the zero-inflated model is a mixture distribution with a probability mass on $[Y=0]$ and a density in $[Y\neq 0]$. It should be noted that when the baseline distribution is either continuous with a pdf $f_{\boldsymbol\theta}(y)$ or discrete but with $p_0(\boldsymbol\theta)=0$, the zero-inflated model is essentially the same as the corresponding zero-altered model.

Commonly used baseline distributions include Gaussian, half-normal, Poisson, negative binomial, beta binomial, etc. The corresponding zero-inflated models are known as zero-inflated Gaussian (ZIG), zero-inflated half-normal (ZIHN), zero-inflated Poisson (ZIP), zero-inflated negative binomial (ZINB), zero-inflated beta binomial (ZIBB), respectively.

\subsection{Maximum likelihood estimate for zero-inflated models}\label{sec:zimle}

Given a random sample $Y_1, \ldots, Y_n$ from the zero-inflated model~\eqref{eq:zimodel}, we adopt the maximum likelihood estimate $\hat\phi$ for $\phi$ and $\hat{\boldsymbol\theta}$ for $\boldsymbol\theta$. Similar as in Section~\ref{sec:hurdle}, we denote $m= \#\{i:Y_i \neq 0\}$.

If the baseline distribution satisfies $P_{\boldsymbol\theta}(Y=0)=0$, then the likelihood function $L(\phi, \boldsymbol\theta) = \phi^{n-m} (1-\phi)^m \cdot \prod_{i: Y_i \neq 0} f_{\boldsymbol\theta}(Y_i)$. Then $\hat\phi = 1-m/n$ and $\hat{\boldsymbol\theta} = {\rm argmax}_{\boldsymbol\theta} \prod_{i: Y_i \neq 0} f_{\boldsymbol\theta}(Y_i)$, which are the same as the ones for hurdle models.

For general cases, the likelihood function for model~\eqref{eq:zimodel} is
\begin{equation}\label{eq:discreteL}
L(\phi, \boldsymbol{\theta}) 
=\left[\phi + (1-\phi)p_0({\boldsymbol\theta}) \right]^{n-m} (1-\phi)^{m}
 \left(1 - p_0(\boldsymbol\theta)\right)^{m} \prod_{i: y_i \neq 0} \rm f_{\rm tr}(Y_{i} \mid \boldsymbol\theta)
\end{equation}
where $f_{\rm tr}(y \mid \boldsymbol\theta) = f_{\boldsymbol\theta}(y)/[1-p_0(\boldsymbol\theta)], y\neq 0$ is the pmf or pdf of the zero-truncated version of the baseline distribution. 
By reparametrization, we let
\begin{equation}
	\psi = 1- [\phi + (1-\phi) p_0({\boldsymbol\theta}) ] = (1-\phi) [1-p_0({\boldsymbol\theta})]
\end{equation}
Then $\phi = 1 - \psi/[1 - p_0(\boldsymbol\theta)]$ and the likelihood of $\psi$ and $\boldsymbol\theta$ is
\[
L(\psi, \boldsymbol\theta)
=(1-\psi)^{n-m}\psi^{m}\cdot \prod_{i: Y_i \neq 0} \rm f_{\rm tr}(Y_i \mid \boldsymbol\theta)
\] 
which is separable for $\psi$ and $\boldsymbol\theta$.

\begin{theorem}\label{thm:zimle}
Let $\boldsymbol\theta_* = {\rm argmax}_{\boldsymbol\theta} \prod_{i: Y_i \neq 0} f_{\rm tr}(Y_i \mid \boldsymbol\theta)$. The maximum likelihood estimate $(\hat\phi, \hat{\boldsymbol\theta})$ maximizing \eqref{eq:discreteL} can be obtained as follows:
\begin{itemize}
    \item[(1)] If $m/n \leq 1-p_0({\boldsymbol\theta_*})$, then $\hat{\boldsymbol\theta} = \boldsymbol\theta_*$ and $\hat\phi = 1 - m/n\cdot (1-p_0({\boldsymbol\theta_*}))^{-1}$.
\item[(2)] Otherwise, $\hat{\boldsymbol\theta} = {\rm argmax}_{\boldsymbol\theta}  (1-\psi(\boldsymbol\theta))^{n-m} \psi(\boldsymbol\theta)^m  \prod_{i: Y_i \neq 0} f_{\rm tr}(Y_i \mid \boldsymbol\theta)$ and $\hat\phi = 1 - \psi(\hat{\boldsymbol\theta}) \cdot (1-p_0(\hat{\boldsymbol\theta}))^{-1}$, where $\psi(\boldsymbol\theta) = \min\{m/n, 1-p_0({\boldsymbol\theta})\}$.
\end{itemize}
\end{theorem}

\medskip\noindent
{\bf Proof of Theorem~\ref{thm:zimle}:}
First of all, we denote 
$\psi_* = {\rm argmax}_{\psi} (1-\psi)^{n-m} \psi^m$ and $\boldsymbol\theta_* = {\rm argmax}_{\boldsymbol\theta} \prod_{i: Y_i \neq 0} f_{\rm tr}(Y_i \mid \boldsymbol\theta)$. It can be verified that $\psi_* = m/n$.

On the other hand, $\psi = (1-\phi) [1-p_0({\boldsymbol\theta})]$ with $\phi \in [0,1]$, which implies $\psi \in [0, 1-p_0({\boldsymbol\theta})]$. If $m/n \leq 1-p_0({\boldsymbol\theta_*})$, then $\hat\psi = m/n$, $\hat{\boldsymbol\theta} = \boldsymbol\theta_*$ is the mle. In this case, the mle of $\phi$ is $\hat\phi = 1 - \hat\psi (1-p_0({\boldsymbol\theta_*}))^{-1}$.

Otherwise, we have $m/n > 1-p_0({\boldsymbol\theta_*})$. Then $\hat\psi = \psi(\boldsymbol\theta) = \min\{m/n, 1-p_0({\boldsymbol\theta})\}$ is the mle of $\phi$ given $\boldsymbol\theta$. In order to find the mle of $\phi$ and $\boldsymbol\theta$, we first find $\boldsymbol\theta^* = {\rm argmax}_{\boldsymbol\theta} L(\psi(\boldsymbol\theta), \boldsymbol\theta)$. Then $\hat{\boldsymbol\theta} = {\boldsymbol\theta}^*$ and $\hat\psi = \psi(\boldsymbol\theta^*)$.
\hfill{$\Box$}

Theorem~\ref{thm:zimle} actually makes the connection between MLEs for zero-inflated models and zero-altered models when $m/n \leq 1 - p_0(\boldsymbol\theta_*)$. It can also be used for finding MLEs numerically.

\begin{example} {\rm\quad
Let $\boldsymbol\theta = (n, \alpha, \beta)$. The pmf of beta binomial distribution is
\[
f_{\boldsymbol\theta} (y) = \dbinom{n}{y} \frac{{\rm beta}(y + \alpha, n - y + \beta)}{{\rm beta}(\alpha,\beta)}
\]
with $y=0, 1, \ldots, n$ and
\begin{eqnarray*}
p_0(\boldsymbol\theta) &=& \frac{\Gamma(n + \beta) \Gamma(\alpha + \beta)}{\Gamma(n + \alpha + \beta)  \Gamma(\beta)}\\
\frac{p_0(\boldsymbol\theta)}{1 - p_0(\boldsymbol\theta)} &=& \frac{\Gamma(n + \beta) \Gamma(\alpha + \beta)}{\Gamma(n + \alpha + \beta) \Gamma(\beta) - \Gamma(n + \beta) \Gamma(\alpha + \beta)}
\end{eqnarray*}
Let $\Psi(\cdot) = \Gamma'(\cdot)/\Gamma(\cdot)$, known as the {\it digamma} function. In order to apply Theorem~\ref{thm:zimle}, we need the following formulas:
\begin{eqnarray*}
\frac{\partial\log f_{\boldsymbol\theta}(y)}{\partial n} &=& \Psi(n+1) - \Psi(n-y+1) + \Psi(n-y+\beta) - \Psi(n+\alpha+\beta)\\
\frac{\partial\log f_{\boldsymbol\theta}(y)}{\partial \alpha} &=& \Psi(y+\alpha) - \Psi(n+\alpha+\beta) + \Psi(\alpha+\beta) - \Psi(\alpha)\\
\frac{\partial\log f_{\boldsymbol\theta}(y)}{\partial \beta} &=& \Psi(n-y+\beta) - \Psi(n+\alpha+\beta) +  \Psi(\alpha+\beta) - \Psi(\beta) \\
\frac{\partial \log p_0(\boldsymbol\theta)}{\partial n} &=& \Psi(n+\beta) - \Psi(n+\alpha+\beta) \\
\frac{\partial \log p_0(\boldsymbol\theta)}{\partial \alpha} &=& \Psi(\alpha+\beta) - \Psi(n+\alpha+\beta) \\
\frac{\partial\log p_0(\boldsymbol\theta)}{\partial \beta} &=& \Psi(n+\beta) + \Psi(\alpha+\beta) - \Psi(n+\alpha+\beta) - \Psi(\beta)
\end{eqnarray*}
}\hfill{$\Box$}
\end{example}

\subsection{Asymptotic properties and Fisher information matrix for zero-inflated MLEs}

Let $Y_1, \ldots, Y_n$ be a random sample from the zero-inflated model~\eqref{eq:zimodel}. The log-likelihood function of $\phi$ and $\boldsymbol\theta$ is
\begin{eqnarray*}
l(\phi, \boldsymbol\theta) = \log L(\phi, \boldsymbol\theta) &&= \log [\phi + (1-\phi) p_0(\boldsymbol\theta)] \cdot \sum_{i=1}^n {\mathbf 1}_{\{Y_i=0\}}
+ \log(1-\phi) \cdot \sum_{i=1}^n {\mathbf 1}_{\{Y_i\neq 0\}} \\
&&+ \sum_{i=1}^n \log f_{\boldsymbol\theta}(Y_i) {\mathbf 1}_{\{Y_i\neq 0\}}
\end{eqnarray*}
Then 
\begin{eqnarray*}
\frac{\partial l}{\partial\phi} &=& \frac{1 - p_0(\boldsymbol\theta)}{\phi + (1-\phi) p_0(\boldsymbol\theta)} \cdot \sum_{i=1}^n {\mathbf 1}_{\{Y_i=0\}} - \frac{1}{1-\phi} \cdot \sum_{i=1}^n {\mathbf 1}_{\{Y_i\neq 0\}}\\
\frac{\partial l}{\partial \boldsymbol\theta} &=& \frac{(1-\phi)p_0(\boldsymbol\theta)}{\phi + (1-\phi) p_0(\boldsymbol\theta)}\cdot \frac{\partial \log p_0(\boldsymbol\theta)}{\partial \boldsymbol\theta} \sum_{i=1}^n {\mathbf 1}_{\{Y_i = 0\}} + \sum_{i=1}^n \frac{\partial \log f_{\boldsymbol\theta}(Y_i)}{\partial \boldsymbol\theta} {\mathbf 1}_{\{Y_i\neq 0\}}
\end{eqnarray*}

\begin{lemma}\label{lem:zimodelmlefirst}
Suppose $0\leq \phi <1$. Then 
\[
E\left(\frac{\partial l}{\partial \phi}\right) = 0 \>\>\mbox{ and }\>\>
E\left(\frac{\partial l}{\partial \boldsymbol\theta}\right) = n(1-\phi) E\left[\frac{\partial \log f_{\boldsymbol\theta}(Y')}{\partial \boldsymbol\theta} \right]
\]
which is $0$ if and only if $E[\partial \log f_{\boldsymbol\theta}(Y')/\partial \boldsymbol\theta]=0$, where $Y'$ follows the baseline distribution $f_{\boldsymbol\theta}(y)$.
\end{lemma}

\medskip\noindent
{\bf Proof of Lemma~\ref{lem:zimodelmlefirst}:}
Since $P(Y_i=0) = \phi + (1-\phi) p_0(\boldsymbol\theta)$ and $P(Y_i \neq 0) = (1-\phi) [1- p_0(\boldsymbol\theta)]$, then $E(\partial l/\partial\phi) = 0$. Let $Y_1', \ldots, Y_n'$ be iid $\sim f_{\boldsymbol\theta} (y)$. Then
\begin{eqnarray*}
E\left[\frac{\partial \log f_{\boldsymbol\theta}(Y_i)}{\partial \boldsymbol\theta} {\mathbf 1}_{\{Y_i\neq 0\}}\right] &&= (1-\phi) \cdot E\left[\frac{\partial \log f_{\boldsymbol\theta}(Y_i')}{\partial \boldsymbol\theta} {\mathbf 1}_{\{Y_i'\neq 0\}}\right] \\ 
&&= (1-\phi) \left\{ E\left[\frac{\partial \log f_{\boldsymbol\theta}(Y_i')}{\partial \boldsymbol\theta} \right] - p_0(\boldsymbol\theta)\cdot \frac{\partial \log p_0(\boldsymbol\theta)}{\partial \boldsymbol\theta}\right\}
\end{eqnarray*}
and
\[
E\left(\frac{\partial l}{\partial \boldsymbol\theta}\right) = (1-\phi) \sum_{i=1}^n E\left[\frac{\partial \log f_{\boldsymbol\theta}(Y_i')}{\partial \boldsymbol\theta} \right] = n(1-\phi) E\left[\frac{\partial \log f_{\boldsymbol\theta}(Y')}{\partial \boldsymbol\theta} \right]
\]
\hfill{$\Box$}


Under regularity conditions, $E[\partial \log f_{\boldsymbol\theta}(Y')/\partial \boldsymbol\theta]=0$ and thus $E\left(\partial l/\partial \boldsymbol\theta\right) =0$ according to Lemma~\ref{lem:zimodelmlefirst}. 

\begin{theorem}\label{thm:zi_Fisher}
Let $Y_1, \ldots, Y_n$ be a random sample from zero-inflated model~\eqref{eq:zimodel}. Under regularity conditions, the Fisher information matrix of the sample is
\[
{\mathbf F}_{\rm ZI} = 
n \left[\begin{array}{cc}
\frac{1-p_0(\boldsymbol\theta)}{[\phi + (1-\phi)p_0(\boldsymbol\theta)] (1-\phi)} & \frac{p_0(\boldsymbol\theta)}{\phi + (1-\phi) p_0(\boldsymbol\theta)} \cdot \frac{\partial \log p_0(\boldsymbol\theta)}{\partial \boldsymbol\theta^T} \\
\frac{p_0(\boldsymbol\theta)}{\phi + (1-\phi) p_0(\boldsymbol\theta)} \cdot \frac{\partial \log p_0(\boldsymbol\theta)}{\partial \boldsymbol\theta}  & {\mathbf F}_{\rm ZI22}
\end{array}\right]
\]
where 
\[
{\mathbf F}_{\rm ZI22} = -(1-\phi) \left(E\left[\frac{\partial^2 \log f_{\boldsymbol\theta}(Y')}{\partial \boldsymbol\theta \partial \boldsymbol\theta^T} \right] + \frac{\phi p_0(\boldsymbol\theta)}{\phi + (1-\phi) p_0(\boldsymbol\theta)} \cdot \frac{\partial \log p_0(\boldsymbol\theta)}{\partial \boldsymbol\theta}\cdot \frac{\partial \log p_0(\boldsymbol\theta)}{\partial \boldsymbol\theta^T}\right)
\]
and $Y'$ follows the baseline distribution $f_{\boldsymbol\theta}(y)$.
\end{theorem}

\medskip\noindent
{\bf Proof of Theorem~\ref{thm:zi_Fisher}:}
Since 
\begin{eqnarray*}
\frac{\partial\log f_{\rm ZI} (y \mid \phi, \boldsymbol\theta)}{\partial \phi} &=& \frac{1-p_0(\boldsymbol\theta)}{\phi + (1-\phi) p_0(\boldsymbol\theta)} {\mathbf 1}_{\{y=0\}} - \frac{1}{1-\phi} {\mathbf 1}_{\{y\neq 0\}}\\
\frac{\partial\log f_{\rm ZI} (y \mid \phi, \boldsymbol\theta)}{\partial \boldsymbol\theta} &=& \frac{(1-\phi) p_0(\boldsymbol\theta)}{\phi + (1-\phi) p_0(\boldsymbol\theta)} \cdot \frac{\partial\log p_0(\boldsymbol\theta)}{\partial \boldsymbol\theta} {\mathbf 1}_{\{y = 0\}} + \frac{\partial \log f_{\boldsymbol\theta}(y)}{\partial \boldsymbol\theta} {\mathbf 1}_{\{y\neq 0\}}
\end{eqnarray*}
Then 
\begin{eqnarray*}
\frac{\partial^2\log f_{\rm ZI} (y \mid \phi, \boldsymbol\theta)}{\partial \phi^2} &=& -\frac{[1-p_0(\boldsymbol\theta)]^2}{[\phi + (1-\phi) p_0(\boldsymbol\theta)]^2} {\mathbf 1}_{\{y=0\}} - \frac{1}{(1-\phi)^2} {\mathbf 1}_{\{y\neq 0\}}\\
\frac{\partial^2\log f_{\rm ZI} (y \mid \phi, \boldsymbol\theta)}{\partial \phi \partial \boldsymbol\theta^T} &=& -\frac{p_0(\boldsymbol\theta)}{[\phi + (1-\phi)p_0(\boldsymbol\theta)]^2} \cdot \frac{\partial \log p_0(\boldsymbol\theta)}{\partial \boldsymbol\theta^T} {\mathbf 1}_{\{y=0\}}\\
\frac{\partial^2\log f_{\rm ZI} (y \mid \phi, \boldsymbol\theta)}{\partial \boldsymbol\theta \partial \phi} &=& -\frac{p_0(\boldsymbol\theta)}{[\phi + (1-\phi)p_0(\boldsymbol\theta)]^2} \cdot \frac{\partial \log p_0(\boldsymbol\theta)}{\partial \boldsymbol\theta} {\mathbf 1}_{\{y=0\}}\\
\frac{\partial^2\log f_{\rm ZI} (y \mid \phi, \boldsymbol\theta)}{\partial \boldsymbol\theta \partial \boldsymbol\theta^T} &=& 
\frac{\phi (1-\phi) p_0(\boldsymbol\theta)}{[\phi + (1-\phi)p_0(\boldsymbol\theta)]^2} \cdot \frac{\partial\log p_0(\boldsymbol\theta)}{\partial \boldsymbol\theta} \cdot \frac{\partial\log p_0(\boldsymbol\theta)}{\partial \boldsymbol\theta^T}{\mathbf 1}_{\{y = 0\}}\\
&&+\frac{(1-\phi) p_0(\boldsymbol\theta)}{\phi + (1-\phi) p_0(\boldsymbol\theta)} \cdot \frac{\partial^2\log p_0(\boldsymbol\theta)}{\partial \boldsymbol\theta \partial \boldsymbol\theta^T} {\mathbf 1}_{\{y = 0\}} +\frac{\partial^2 \log f_{\boldsymbol\theta}(y)}{\partial \boldsymbol\theta \partial \boldsymbol\theta^T} {\mathbf 1}_{\{y\neq 0\}}
\end{eqnarray*}
The rest of the conclusions can be obtained via $l(\phi, \boldsymbol\theta) = \sum_{i=1}^n \log f_{\rm ZI} (Y_i \mid \phi, \boldsymbol\theta)$, $P(Y_i = 0) = \phi + (1-\phi)p_0(\boldsymbol\theta)$ and $P(Y_i \neq 0) = (1-\phi) [1-p_0(\boldsymbol\theta)]$.
\hfill{$\Box$}

As direct conclusions of Theorem~18 in \cite{ferguson1996}, under regularity conditions, we have (i) 
\[
\sqrt{n}(\hat\phi - \phi_0)  \stackrel{{\cal L}}{\longrightarrow} N\left(0, \frac{[\phi + (1-\phi)p_0(\boldsymbol\theta)] (1-\phi)}{1-p_0(\boldsymbol\theta)}\right)
\] 
and (ii) $\sqrt{n}(\hat{\boldsymbol\theta} - \boldsymbol\theta_0)  \stackrel{{\cal L}}{\rightarrow} N\left(0, {\mathbf F}_{\rm ZI22}^{-1} \right)$. However, $\hat\phi$ and $\hat{\boldsymbol\theta}$ are usually not asymptotically independent unless $p_0(\boldsymbol\theta) = 0$ or does not depend on $\theta$. These results can be used for building up confidence intervals of $\phi$ and $\boldsymbol\theta$ as well.

\begin{example}\label{ex:zipFisher} {\rm
For zero-inflated Poisson (ZIP) model, the pmf of the baseline distribution is $f_\lambda(y)= e^{-\lambda}\lambda^{y}/y!$ with $y=0, 1, \ldots$, where $p_0(\lambda)= e^{-\lambda}$. Then $\partial \log f_\lambda(y)/\partial \lambda = y/\lambda -1$.

According to Theorem~\ref{thm:zi_Fisher}, the Fisher information matrix of the Poisson sample is
\[
{\mathbf F}_{\rm ZIP} = 
n \left[\begin{array}{cc}
\frac{1-e^{-\lambda}}{[\phi + (1-\phi)e^{-\lambda}] (1-\phi)} & -\frac{e^{-\lambda}}{\phi + (1-\phi) e^{-\lambda}} \\
-\frac{e^{-\lambda}}{\phi + (1-\phi) e^{-\lambda}}  &  (1-\phi) \left(\frac{1}{\lambda}  - \frac{\phi e^{-\lambda}}{\phi + (1-\phi) e^{-\lambda}} \right)
\end{array}\right]
\]
Note that $\frac{1}{\lambda}  - \frac{\phi e^{-\lambda}}{\phi + (1-\phi) e^{-\lambda}} > 0$ as long as $\lambda > 0$ and $0\leq \phi < 1$.
\hfill{$\Box$}
}\end{example}


\section{Microbiome Data Application}
As an application, the bootstrap KS test with unknown parameters (Algorithm 1) has been used to a list of 229 bacterial and fungal OTUs~\citep{aldirawi2019identifying,tipton2018fungi}. We are interested in knowing how many of the 229 OTU follows each of the following distributions given that the distribution parameters are unknown: Poisson, negative binomial, beta binomial, beta negative binomials, and the corresponding zero-inflated and hurdle models. Table~\ref{ks}, which was regenerated  from \cite{aldirawi2019identifying}, summarizes the number of features that do
not show significant divergence (p-value $>$ 0.05).

Poisson, negative binomial, ZIP, ZINB have been used commonly for modeling microbiome data. However, as shown in the above table, Poisson, zero-inflated Poisson, and Poisson hurdle are not appropriate distributions to model sparse microbial features as only 0.4\%, 2\%, 1\% out of 229 the features were able to be appropriately fitted using these distributions, respectively. On the other hand, binomial and negative binomial families can be used to approximate sparse microbial data, with BNBH as the best distribution to model such dataset (being able to appropriately fit 53\% of the 229 features) using the proposed conservative method. In addition, ZIBNB fits about half of the features followed by BBH and ZIBB which they fit about 40\% of the features.

Based on the above table, we conclude that zero-inflated and hurdle beta binomial or beta negative binomial are more appropriate than commonly used models.

\begin{table}[]
\caption{Number and percentage of species out of 229 species that don't show significant difference (KS p-value $>0.05$)}
\label{ks}
\begin{center}
\resizebox{0.85\textwidth}{!}{
\begin{tabular}{|| c| c |c||}
\hline
     \textbf{Distribution} & \textbf{Number} & \textbf{Percentage}    \\
     \cline{1-3}
     Poisson & $1 $ & 0.4\%  \\
     \hline
     negative binomial (NB) & 23 & 10\%    \\
     \hline
     beta binomial (BB) & 76 & 33\%   \\
     \hline
     beta negative binomial (BNB) &  60 & 26\%  \\
     \hline
     zero-inflated Poisson (ZIP)&  3 & 2\%         \\
     \hline
      zero-inflated negative binomial (ZINB)&  25 &  11\%         \\
     \hline
      zero-inflated beta binomial (ZIBB)&   89 & 39\%        \\
     \hline
      zero-inflated beta negative binomial (ZIBNB)&  110& 48\%         \\
     \hline
     Poisson hurdle (PH)&  2 & 1\%       \\
     \hline
     negative binomial hurdle (NBH)& 56 & 24\%           \\
     \hline
     beta binomial hurdle (BBH)&  92 & 40\%        \\
     \hline
     beta negative binomial hurdle (BNBH)& 121 & 53\%          \\
     \hline
 \end{tabular}
\label{tab1}
}
\end{center}
\end{table}

\section{Conclusion}\label{conclusion}

Understanding the role of the microbiome in human health and how it can be modeled is becoming increasingly relevant for preventive medicine and for the medical management of chronic diseases \citep{calle2019statistical}. However, the microbiome data is highly sparse and skewed. It is very challenging to select an appropriate probabilistic model. 

In this paper, we use the MLE approach to estimate the parameters of general zero-inflated and hurdle models. We also derive the corresponding Fisher information matrices for exploring the estimator's asymptotic properties to build up the confidence intervals of the parameters.

In the literature, Poisson and negative binomial models have been commonly used for modeling microbiome data. Based on a real dataset analysis, we show that zero-inflated (or hurdle) beta binomial or beta negative binomial models are more appropriate.

\section*{Acknowledgments}

This work was partly supported by NSF grant DMS-1924859.

\end{document}